\documentclass[preprint]{vgtc}               

\ifpdf
  \pdfoutput=1\relax                   
  \usepackage{graphicx}                
  \DeclareGraphicsExtensions{.pdf,.png,.jpg,.jpeg} 
\else
  \usepackage{graphicx}                
  \DeclareGraphicsExtensions{.eps}     
\fi%

\usepackage{subfigure}

\graphicspath{{figs/}{pictures/}{images/}{./}} 

\usepackage{microtype}                 
\PassOptionsToPackage{warn}{textcomp}  
\usepackage{textcomp}                  
\usepackage{mathptmx}                  
\usepackage{times}                     
\usepackage{cite}                      
\usepackage{tabu}                      
\usepackage{booktabs}                  
\usepackage{enumitem}

\onlineid{0}

\vgtccategory{Research}

\vgtcinsertpkg

\title{Show Me My Users: A Dashboard Visualizing User Interaction Logs}

\author{Jinrui Wang\thanks{Both authors contributed equally.} \thanks{e-mail: j.wang-245@sms.ed.ac.uk}\\ %
        \scriptsize The University of Edinburgh %
\and Mashael AlKadi*\thanks{e-mail: malkadi@iau.edu.sa}\\ %
     \parbox{1.4in}{\scriptsize \centering The University of Edinburgh \\ Imam Abdulrahman bin Faisal University} %
\and Benjamin Bach\thanks{e-mail: bbach@exseed.ed.ac.uk}\\ %
     \scriptsize The University of Edinburgh}

\abstract{%
This paper describes the design of a dashboard and analysis pipeline to monitor users of visualization tools in the wild. Our pipeline describes how to extract analytical KPIs from extensive log event data involving a mix of user types. The resulting three-page dashboard displays live KPIs, helping analysts understand users, detect exploratory behaviors, plan education interventions, and improve tool features. We propose this case study as a motivation to use the dashboard approach for a more `casual' monitoring of users and building carer mindsets for visualization tools.
}

\teaser{
  \centering
  \includegraphics[width=0.9\linewidth]{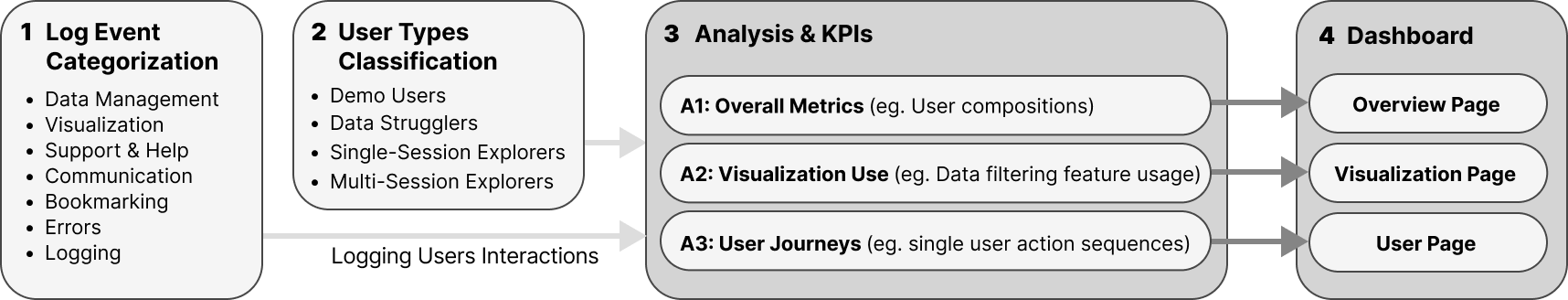}
  \caption{%
  	\textbf{Dashboard Design Pipeline of Visualization Tool Logs}: User interactions are logged and categorized by event types (1) to determine different stages of visual exploration. With user types classified (2), we are then able to apply analysis and calculations (3) according to our key performance indicators (KPIs). The dashboard (4) pages and contents are informed directly by the KPIs.
  }
  \label{fig:teaser}
}

\CCScatlist{
  \CCScatTwelve{Human-centered computing}{Visu\-al\-iza\-tion}{Visualization design and evaluation methods}{};
  \CCScatTwelve{Human-centered computing}{Visu\-al\-iza\-tion}{Visu\-al\-iza\-tion application domains}{Visual Analytics};
}

\begin{document}


\firstsection{Introduction}

\maketitle

Understanding how people use a visualization tool in their daily work environment is key to improve tool features and provide users with adequate training. In particular, we are interested in understanding where people struggle to use a given visualization design or interactive features 
over time~\cite{bach2023challenges} and informing decisions about specific features, identifying users in need for help, developing new training methods, or planning for in-person interventions (e.g., workshops, drop-in sessions). 
This goal poses two distinct yet intertwined questions: \textit{(Q1) What information to capture about users using the tool? and 
(Q2) How to present these data to an analyst to support decision-making?}

Current methods to collect and analyze usage data across a diverse crowd of users (anonymous, geographically distributed, with different backgrounds, tasks, and data) over a longer period of time do not support these tasks well. For example, mini-questionnaires~\cite{molineroUnderstandingUseVistorian2017} that prompt users for written feedback or online check-in session~\cite{alkadi2022understanding} can respond to a user's specific questions but require active engagement by users, qualitative analysis, and do hardly scale. Interaction logs do capture users' interaction events with the tool but their analysis usually requires bespoke analysis scripts~\cite{von2014interaction,gathani2022grammar,InsightMetrics,rae2022understanding,sukumar2020characterizing} or time-intensive manual exploration~\cite{monroe2013temporal}. 

In this paper, we design a dashboard and analysis pipeline that provides an at-a-glance live-view to monitor user types and their engagement with the visualization software. The approach is inspired by the use of analytics and dashboards in the field of Learning Analytics~\cite{lemo_LA,khosravi2021intelligent}. However, unlike those approaches, which track students' progress throughout an often well-defined course structure, visualization systems are much more open; using multiple visualization designs and interactive features, loading individual data sets, and developing personal exploration and analysis strategies. Hence, our pipeline (\autoref{fig:teaser}) features a range of interaction events, their collection mechanism, classifying users into different types, calculating key-performance indicators (KPIs) on the collected data, and designing an interactive visual dashboard with three pages. The resulting live-overview allows to understand general trends in tool use, identify the impact of changes to features, training resources or educational activities, and plan for future interventions. 

We base our design on a case study with \textit{The Vistorian} (\url{vistorian.net})~\cite{alkadi2022understanding,bach2015networkcube}, a research prototype platform for delivering interactive network visualizations to a large set of end users. The dashboard was developed over several rounds of iterations among the authors of this paper, who at the same time take the role of the data analyst to inform their research around the Vistorian. In the following, we describe our analysis pipeline (\autoref{sec:analysis}) and dashboard design (\autoref{sec:dashboard}), then discuss our work through findings from up to 2298 users using the Vistorian over 28 months (\autoref{sec:findings}). We close by discussing implications and lessons learned from the dashboard and the potential of monitoring user behavior in visualization tools live (\autoref{sec:discussion}). A live version of the dashboard is online via \url{https://halcyonwjr.github.io/Vistorian_Dashboard/} with more details in github repository \footnote{\url{https://github.com/HalcyonWjr/Vistorian_Dashboard}} and the supplementary material.

\section{Background and Related Work}

\paragraph{\textbf{The Vistorian Use Case}}
The Vistorian is an open-source network visualization web application with five interactive visualizations: node-link diagram, adjacency matrix, timeline, map, and a coordinated view combining node-link and matrix. It supports users in uploading data, creating networks, and exploring visualizations interactively with various help resources such as examples or visualization manuals. The description of different help resources is included in the supplemental material (Appendix C). The Vistorian has been used in numerous workshops at scientific conferences, bespoke network visualization workshops, and summer schools~\cite{alkadi2022understanding}. Its main target audience is analysts without programming skills. While the authors of this paper have received plenty of informal feedback from workshops about how people use the tool, we have not had any means to monitor progress during and outside of workshops.

\paragraph{\textbf{Interaction Log Analysis}}
Interaction logs capture single interaction events and are commonly used for webpages and software systems. In visualization research, they play a critical role in many studies to describe user intentions, behaviors\cite{battleCharacterizingExploratoryVisual2019}, provenance\cite{pachuiloLeveragingInteractionHistory2016}, and workflows either in different levels of granularity \cite{gathani2022grammar} or as a sequence \cite{InsightMetrics}. Many of these case studies are based on small datasets obtained from lab settings \cite{gathani2022grammar} with clear analysis tasks\cite{blascheckLoggingInteractionsLearn2016}. A common way to visualize user interaction events is through timelines~\cite{monroe2013temporal} or diagrams (e.g., \cite{zgraggen2015eries}).

The field of Learning Analytics focuses on investigating users interactions with the goal to understand people's learning and progression through digital learning platforms~\cite{beheshitha2016role,matcha2019analytics,mutlu2021towards}, educational videos \cite{li2015mooc}, and academic performance through courses and learning management systems \cite{matcha2019systematic,matcha2019analytics,echeverria2018exploratory,kuosa2016interactive}. Our work builds on these approaches and extends them to the open nature of interactive multi-view visualization systems. These systems lack explicit assessments of users' progress such as tests and assessments.

\paragraph{\textbf{Dashboards design}}

While dashboards are used across many applications, including learning analytics~\cite{matcha2019systematic,duval2011attention}, designing dashboards is not trivial. It requires contemplating the analysts' tasks and developing solutions that make the best use of the available screen space and visualization affordances.
Sarikaya et al. \cite{sarikayaWhatWeTalk2019a} described example dashboards by their goals, purposes, audiences, visual features, and data semantics, among which an in-depth analytic dashboard requires careful choice of data, visual layout, as well as high demand on designer and user's visual literacy. More actionable, Bach et al.~\cite{bachDashboardDesignPatterns2023} described design patterns for dashboards, e.g., layouts, visualizations, or structures. Qu et al.~\cite{quKeepingMultipleViews2018} studied how designers balance visual consistency across multiple views, and Aparício developed a BI dashboard decision tree to support the design process \cite{costa2019supporting}. Our design is based on this guidance and can generalize to visualization systems beyond our use case.

\section{Analysis} 
\label{sec:analysis}

The analysis pipeline consists of the following steps (\autoref{fig:teaser}): Firstly, log events are used to distinguish various user interactions. The log categories (\autoref{fig:teaser}-1) are defined on the basis of the common tasks in the visual exploration process. Then we categorize four user types (\autoref{fig:teaser}-2) based on key factors separating them during exploration stages. From logs and user types, we identify three analytical KPIs with detailed requirements (\autoref{fig:teaser}-3). The calculation outcomes are used to feed the dashboard composition and layout design (\autoref{fig:teaser}-4).

\paragraph{\textbf{Log Events Types and Logs Collection}}

We log 94 types of interaction events using the \textit{Intertrace} JS library\cite{intertrace}. Events are grouped into seven categories: 
\textit{Data-Management} (e.g., create a network), 
\textit{Visualization \& Interaction} (e.g., change visual encodings), 
\textit{Support \& Help} events (e.g., usage of help resources), \textit{Communication} events (e.g., consult the team), \textit{Bookmark} events (e.g., capture/annotate a network state), \textit{Error-Tracking} events (e.g., report issues), and \textit{Activity-Logs} events (e.g., share feedback). The full list of of events is included in the supplemental material (Appendix A). Once a user visits the Vistorian, a unique session ID is assigned to the user's interaction logs. The session ID will last with the user until the browser's cache is cleared or the browser page in incognito mode is closed. In such cases, a new session ID is assigned to the same user after revisiting the Vistorian page. We combine session files if they share the same IP address, and the new session was created within 1 minute or less from the last session's end time (the time to refresh the browser memory and continue using the Vistorian).

\paragraph{\textbf{User Types and Log Calculations}}

Our pipeline features four user types based on 
(i) whether they used their \textbf{own data} or demo data, 
(ii) whether they \textbf{created any own networks}, and 
(iii) whether they \textbf{returned for multiple visits}. 

\vspace{-6pt}
\begin{itemize}[noitemsep, leftmargin=*]
\item \textbf{Demo user:} no own data, but only use the provided demo data.
\item \textbf{Data strugglers:} users who attempted but did not manage to create their own network from a data table. 
\item \textbf{Single-session explorers:} users with a single  session using the Vistorian with their own data.
\item \textbf{Multi-session explorers:} users who return for at least one additional session, at least 20 minutes after their first session. 
\end{itemize}
\vspace{-6pt}

Combining log event types and user types in the logs analysis process assists in calculating various metrics. For example, exploration patterns can be classified based on their themes (e.g. temporal, geographical,..). Also, tool's usability can be measured for each user type instead of just the overall usage. This enables a better understanding of the tool's design and how it supports different users needs. This can be reached by observing the variation in the number of user types over time, examining user decisions through exploration sessions and assessing the impact of educational efforts (e.g. courses).  

\paragraph{\textbf{Requirements and KPIs}} From log event categories and four types of users, we are mostly interested in unfolding KPIs from three aspects: 

\vspace{-6pt}
\begin{itemize}[noitemsep, leftmargin=*]
    \item \textbf{A1: General metrics} about number of users, time spent with the tool, change over time, types of users, etc. 
    \item \textbf{A2: Visualization use} such as time spent with each visualization, trends over time, and use of interactive features. 
    \item \textbf{A3: User journeys} for specific users and their use of the system.
\end{itemize}
\vspace{-6pt}

The specific requirements are listed in (\autoref{tab:dashboard_tasks}), which have been systematically identified in our previous studies on understanding users through a (tedious manual) log analysis alongside interviews, workshops, and drop-in sessions~\cite{alkadi2022understanding}.

\begin{table}[hb]
  \caption{%
        Analytical KPIs list covering three main aspects guiding dashboard design. Codes for each KPI (e.g., a1) refer to specific charts in dashboard.%
    }
  \label{tab:dashboard_tasks}
  \scriptsize%
  \centering%
    \begin{tabu}{p{0.05cm}p{6.8cm}}
  	\toprule
        \multicolumn{2}{l}{\textbf{A1: Overview page (\autoref{fig:page_overview})}}\\
  	1 & Number of total users (a1)\\
        2 & Trend of user visits over time (a2)\\
        3 & Distribution of session length by user type (a3)\\
        4 & Users returning rate (a4)\\
        5 & User visits changing with the external events (b)\\
        6 & Most frequently visited features (c1)\\
        7 & Use of help resources (c2)\\
        8 & Time spent by user type (c3)\\
  	
  	\midrule               
  	\multicolumn{2}{l}{\textbf{A2: Visualization page (\autoref{fig:page_vis}(a))}}\\
        1 & Time spent on each visualization per user (d2)\\
        2 & Number of users using a visualization (d1)\\
        3 & Use of interactive visualization features (d3, d4)\\
        \midrule               
  	\multicolumn{2}{l}{\textbf{A3: User page (\autoref{fig:page_vis}(b))}}\\
        1 & Interactions per individual (e2, e3)\\
        2 & Number of networks successfully created by each user\\
        3 & Exploration sequence for different user types (e1)\\
  	\bottomrule
  \end{tabu}%
\end{table}
\section{Dashboard Design}
\label{sec:dashboard}

\begin{figure*}[t]
  \centering
  \includegraphics[width=1\textwidth]{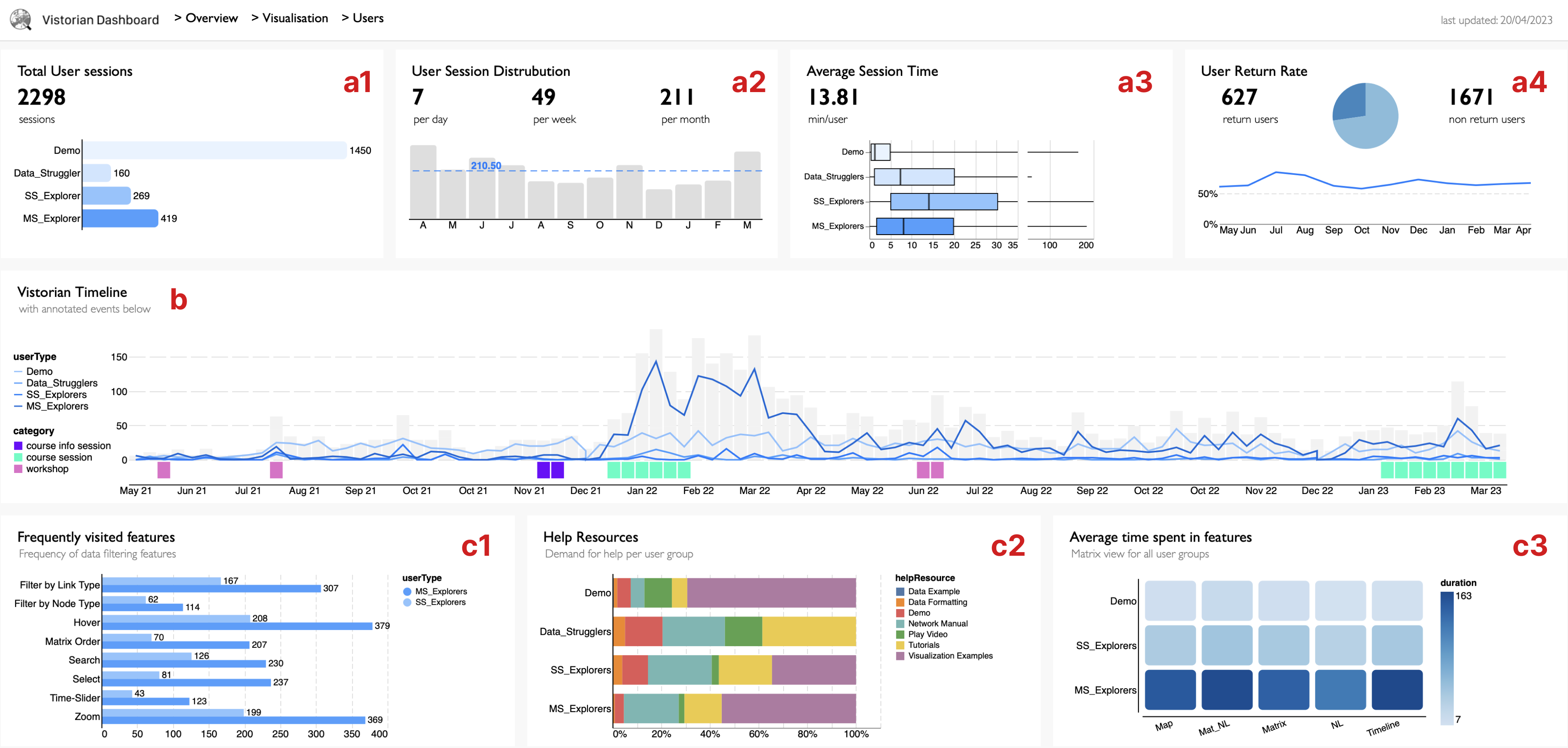}
  \caption{%
  	Page \textbf{Overview}: A navigation bar on top, three stratified rows to report overall trends and usage in the Vistorian.%
  }
  \label{fig:page_overview}
\end{figure*}

The dashboard was designed over a period of 3 months, featuring regular consultation among all the authors. We followed Bach et al's ~\cite{bachDashboardDesignPatterns2023} six-stage design process for dashboards through identifying data abstractions, data board structure and pages, page layout, visualization design, screenspace optimization, and designing interaction.\footnote{\url{https://dashboarddesignpatterns.github.io/processguidelines.html}} Where appropriate, we refer to Bach et al.'s design patterns in \textit{italic} formatting. Throughout the design process, three authors evaluated if the design answers the tasks in (\autoref{tab:dashboard_tasks}), and observed user patterns across charts and pages.

\begin{figure}
  \includegraphics[width=1\columnwidth]{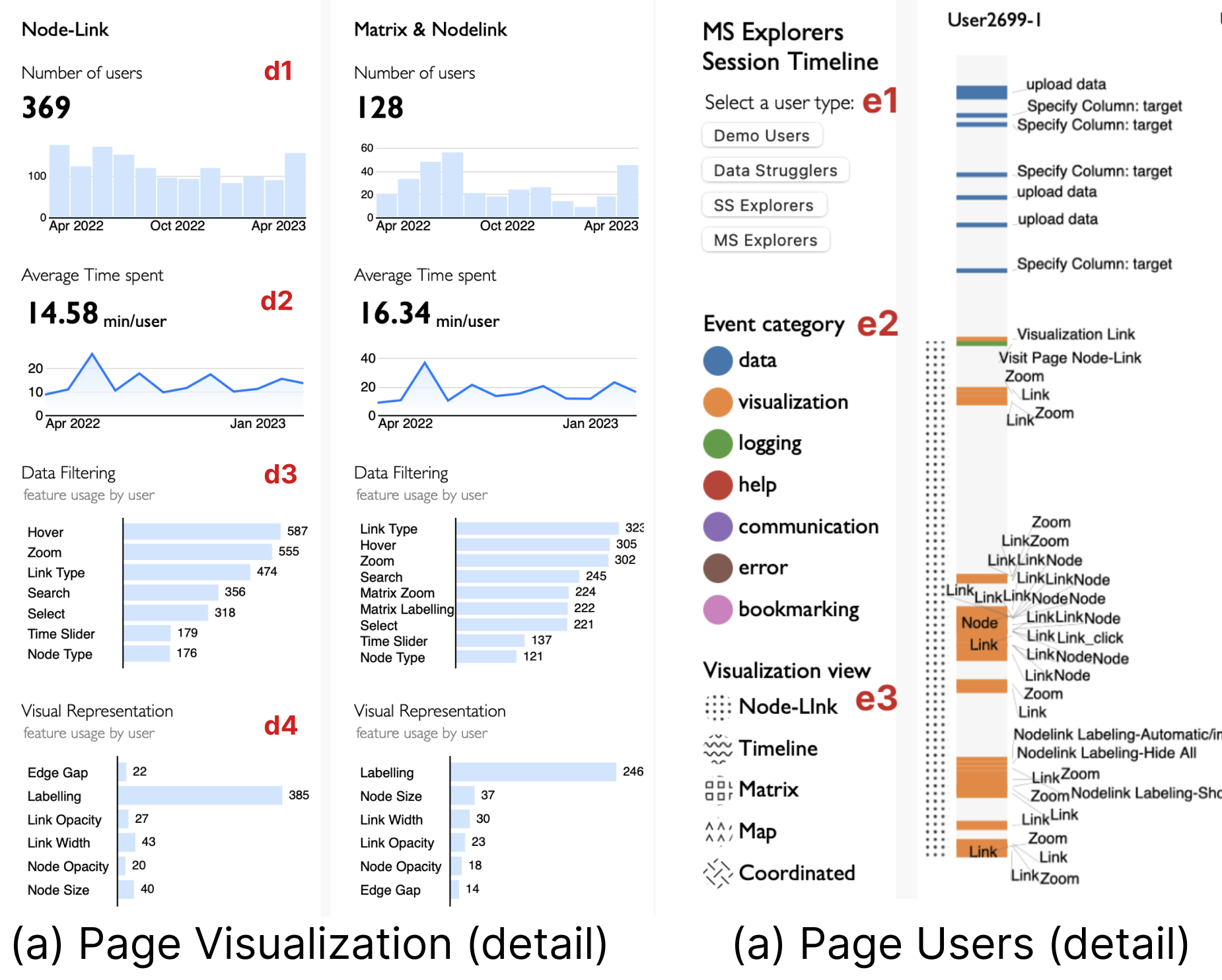}
    \vspace{-2em}
\caption{(a) Page \textbf{Visualization}: a table layout showing data about each visualization. (b) Page \textbf{User}: individual interaction sequences color-coded by event categories. 
}
\label{fig:page_vis}
\vspace{-1em}
\end{figure}

The three main goals (A1, A2, A3) informed a \textit{multi-page structure}. The \textbf{Overview (A1)} page is the \textit{hierarchical} leading page (\autoref{fig:page_overview}), followed by two pages about \textbf{visualization (A2)} (\autoref{fig:page_vis}(a)) and \textbf{users journeys (A3)} (\autoref{fig:page_vis}(b)). Each page is kept in a \textit{screen-fit} size to provide all the information at a glance. Full screenshots of the dashboard can be found in supplementary materials (Appendix B).

\paragraph{\textbf{Overview page}}
The overview page reports on the overall activities with the Vistorian. The \textit{stratified layout} organizes three levels of information as (\autoref{fig:page_overview}) shows: (a1-a4) user statistics, (b) temporal trends, and (c1-c3) key features usage.

To glimpse the initial profile of tool users, the first group of charts speaks of up-to-date facts about user composition (a1), monthly sessions (a2), session length (a3), and retention rates (a4), establishing the context for tool performance. Meanwhile, we are conducting research with the tool about visualization in practice, where we require simple, real-time access to how these efforts impact users, and to compare temporal trends of past efforts. The annotated timeline (b) accommodates this need, allowing us to account for workshops, and training courses in relation to the changes in user numbers. Furthermore, we would like to learn which are the go-to resources for help and whether different types of users show differences in feature preferences. This set of charts answers these questions from: popularity of interaction features (c1), supporting materials (c2), and users’ interest in network visualization by session time (c3).

\paragraph{\textbf{Visualization page}}
The Visualization page uses a \textit{table layout} (\autoref{fig:page_vis}(d)) where each column depicts one visualization view, each row measures shared metric across different views.
Our previous study \cite{alkadi2022understanding} revealed that users differ in view preferences and interaction techniques. Therefore this page includes (d1) the number of users, (d2) time spent on each view, and (d3) popularity in interaction features (\autoref{tab:dashboard_tasks}) to compare different visualizations in use.

We adopted Gotz' and Wen's \cite{gotz2009characterizing} taxonomy of user's intent to classify interactions. The first group includes data filtering (d3), e.g. time slider, filter by node and link type. The second group changes the visual representation (d4) of elements, i.e. label ordering techniques in the matrix view. The dashboard lists the frequency of events triggered to represent the usage popularity. These comparisons also encourage us to reflect on feature and user KPIs without distinguishing views, provoking further speculations into specific trends that are inconsistent with other views or the overall trend.

\paragraph{\textbf{User page}}
Guided by the third aspect (A3), we designed a timeline to show users' action sequences. Previous studies summarized an 8-step visual exploration framework~\cite{alkadi2022understanding}, where actions in each step are traced in our log list. Though the framework can guide the learning process, it will be heavily interrupted and tweaked in practice, where timelines provide an accurate representation.

The user page in \autoref{fig:page_vis}(b) adopts a \textit{grouped layout} where each timeline shows interaction events for an individual user (\autoref{fig:page_vis}(b) showing only one timeline). Each interaction event is visualized as a horizontal dash on the grey timeline, color-coded by its category (e2) while the specific event name is shown as a label. Events that are triggered continuously within a one-second (e.g., continuous hovering of nodes while browsing labels) interval are combined into a block on the timeline and measured using the first to last event timestamp. Adjacent to each session timeline are sidebars indicating the user's current visualization view through different textures (e3). Interactive filtering can be used to show and hide individual types of events (e2).

\section{Findings from Dashboard in Use}
\label{sec:findings}

Over 28 months, we collected logs from 2298 users. Besides users using the Vistorian at home, the time period included a 6-week course on network visualization and the Vistorian,\footnote{\url{https://vistorian.github.io/courses}} a 2-month coaching program with individual drop-in sessions,\footnote{\url{https://vistorian.github.io/upcoming_courses}} and four workshops stretching a few hours each. Below, we report on both findings from the logs and how these findings influenced decisions around the design of tool features and educational efforts.

\textbf{Understanding user trends:} Over 50\% of all users were demo users and just about 5\% (160) were data strugglers (\autoref{fig:page_overview}a1). About 25\% of users returned to the Vistorian for additional sessions (\autoref{fig:page_overview}a4). Those returning (MS\_Explorers) spent an average of 15 minutes on their sessions (\autoref{fig:page_overview}a3) while demo users spent only about 5 min with the tool in each session. From the timeline (\autoref{fig:page_overview}b), we can see a constant, yet slightly increasing number of users (grey bars) with an increase in sessions during the 6-week course (cyan bars under the timeline). During the same period, we can observe a robust increasing trend in the matrix and node-link coordinated view (\autoref{fig:page_vis}), justifying our educational effort. 
In (\autoref{fig:page_overview}c2), we can see that most users use visualization examples (purple), except data strugglers who rely heavily on tutorials (yellow), and videos (green). The more experienced a user becomes (MS\_Explorer), the less they rely on demos (red) and data formatting (orange) help. Analysis of help resources usage leads us to create more help videos and provide more example data sets, especially to support those data strugglers. Likewise, we used the individual timelines (\autoref{fig:page_vis}b) to track users to prepare for our encounters with them in the drop-in sessions.

\textbf{Understanding individual exploration}: Analysis of feature usage provides insights about the \textit{exploratory behavior} per user type and raises further research questions as in (\autoref{fig:page_overview}c1). For example, on average users took about two minutes to upload data and create a network. Those taking longer to create networks without exploration require our attention. We also observed in the individual timelines that most users start by using node-link diagrams, then explore the other visualizations. Currently, we are working on guidelines for self-directed learning to guide users through the entire process of visual network exploration. These observations could lead to new KPIs (e.g., flag users that upload more than 3 data sets without exploration) to inform educators. Likewise, these observations can lead to new hypotheses and questions about, for example, why single-session users explore fewer ordering algorithms in the adjacency matrix, why they tend to explore temporal changes in their networks to a lesser extent, or whether this is related to a user's tasks (e.g., no interest in temporal exploration) or to their skills (e.g., lack of awareness of temporal exploration).

\textbf{Improving tool feature:} Based on the data about data strugglers and our observations from the workshops, we designed and implemented a more comprehensive data uploading wizard with a linear set of configuration steps and detailed explanations at each step~\cite{alkadi2022understanding}. In addition, further coordinated views can be created based on users' common exploration patterns. Currently, we are in the process of adding more help resources, improving interaction affordances, and increasing the range of visualization techniques.
\section{Discussion \& Future Work}
\label{sec:discussion}

In this paper, we present a dashboard design and analysis pipeline for visualizing user interactions with an interactive visualization tool. Our observations show how monitoring users over a period of time can inform decisions for interventions and further analysis. They also 
call for \textbf{more `casual' monitoring of users}, whether the respective tools are research prototypes or mature tools. By \textit{casual}, we mean the \textit{constant, exhaustive, and up-to-date display of interaction log data}. This approach is different from dedicated log analyses performed with either specific questions in mind, focusing on specific aspects of a tool, or not running alongside tool development and education. Our user types can be seen as personas, further informing help resources for particular audiences.

We believe our dashboard design and KPIs are \textbf{generalizable to many other interactive visualization applications}. Adaptions would require extending or limiting the types of events and setting different parameters and thresholds for our user categorization. New KPIs could be added and more widgets and pages designed. For example, we can imagine pages dedicated to interaction features, specific user groups, or specific workshops. We can also imagine a general protocol and architecture for visualization tool logging and visualization with dashboards. Such a framework could also try to feed on \textbf{additional data sources}, such as textual notes from users~\cite{block2020micro}, call information, email exchange, or workshop attendances. However, qualitative data requires bespoke analysis and careful evaluation to what extent the derived KPIs really reflect users' concerns and progress. We need more research to enable KPIs to represent causation rather than correlation.

In particular, we want to encourage the creation of specific \textbf{KPIs that can capture learning or failure in visualization use}, e.g., an increase in the frequency of using a visualization, frequency of help resources. Good KPIs must be able to capture such specific developments, but they are naturally harder to define and require proper evaluation. Such KPIs could feed into \textbf{automatic recommendation systems personalizing help and support} to remote users by avoiding the clippy-fallacy. In the end, we might be able to create individual user profiles and help individualize as well as tailor (i.e., for specific groups of users) features and education. We could finally be able to track learning, i.e., how individual users learn visualization interfaces as well as visualization knowledge. Currently, we still know too little about our users or describe them based on assumptions and informal evidence (e.g., novices, domain experts, data scientists, etc.)~\cite{burns2023we,kauer2021public}. 

However, it is important is to \textbf{create a `carer mindset'} with principles that care about our tools, their deployment and use, their continuous development, and the education to help people master new tools and interfaces. The visualization community is full of stellar examples of care mindsets, e.g., for tools such as D3, Vega, or Eventflow\cite{monroe2013temporal}. We believe, our approach is a step towards building care mindsets for visualization tools and helping increase the impact of data visualization.

\acknowledgments{This research was partly supported by the EPSRC grant EP/V010662/1.}

\bibliographystyle{abbrv-doi}

\bibliography{main}
\end{document}